\documentclass{pazh}
\usepackage{graphicx}
\usepackage{latexsym}

\newcommand{\km}{\,\mbox{km}\,\mbox{s}^{-1}}

\def\farcm{\hbox{$.\mkern-4mu^\prime$}}

\def\SCORPIO{\mbox{SCORPIO}\,}

\begin{document}

\title{The \SCORPIO Universal Focal Reducer of the 6-m Telescope}

\author{ Afanasiev V.L. \and  Moiseev A.V.}

\institute{Special Astrophysical Observatory, RAS, Nizhnii Arkhyz,
Karachai-Cherkessian Republic, 357147 Russia }

\date{ }

\offprints{Alexei Moiseev, \email{moisav@sao.ru}}

\titlerunning{The \SCORPIO focal reducer}
\authorrunning{ Afanasiev \& Moiseev}

\abstract{ We describe the \SCORPIO focal reducer that has been used since the
fall of 2000 for observations on the 6-m Special Astrophysical Observatory
telescope. We give parameters of the instrument in various observing modes
(direct images, long-slit and multislit spectroscopy, spectropolarimetry,
Fabry-Perot panoramic spectroscopy). Observations of various astronomical objects
are used as examples to demonstrate the \SCORPIO capabilities.} \maketitle

\section{INTRODUCTION}

The idea of using a focal reducer on a telescope was suggested and implemented
by Georg Court\'es  as the 50-60 of the last century (see, for example,
Court\'es, 1960, 1964). The focal reducer allows several problems to be solved
simultaneously.  Firstly, the equivalent focal ratio of the system becomes
faster and the field of view increases, which is important for studying faint
extended objects. Secondly, the off-axis aberrations of the primary mirror can
be corrected by using specially calculated optics. Thirdly, it becomes possible
to install  dispersing elements (grisms, Fabry-Perot interferometer (FPI)
etc.) in the parallel beam between the collimator and the camera, which turns
the focal reducer into a universal spectrograph.

Instruments based on this scheme, which are primarily designed for the
spectroscopy and photometry of faint extended objects, have gained wide
acceptance in the last two decades. It will suce to mention the EFOSC camera
of the 3.6-m ESO telescope (Buzzoni et al. 1984); in many respects, it became
the prototype of modern spectrographs for 8-10-m telescopes, such as FORS on
VLT (Nicklas et al. 1997). The first focal reducer for interferometric
observations on the 6-m BTA telescope was created at the Special Astrophysical
Observatory (SAO) of the Russian Academy of Sciences in the mid-1980s using
commercially available photographic lenses. Despite such shortcomings as poor
image quality at the edge of the field of view, low optical transmittance
(about 30\% at maximum), and the absence of any automation, the reducer had
been used on the 6-m BTA telescope for more than ten years until the question
of its upgrading arose. In 1999, work on the  creation of a new focal reducer
for the prime focus of the 6-m BTA telescope began at the SAO. The new \SCORPIO
(Spectral Camera with Optical Reducer for Photometric and Interferometric
Observations) focal reducer has allowed the following types of observations of
extended and starlike objects to be performed at the prime focus of the 6-m
telescope:

\begin{itemize}
\item  Direct images  in broad-, medium- and narrow-band filters.
\item  Panoramic spectroscopy with the FPI.
\item  Long-slit spectroscopy.
\item  Slitless spectroscopy.
\item  Multiobject spectroscopy with 16 slits moved remotely in the focal plane.
\item  Polarimetry in the filtres and spectropolarimetry.
\end{itemize}

The mechanical and optical parts of \SCORPIO were produced at the
SAO breadboard workshops. The first BTA observations were
performed in September 2000 with the old version of the
prime-focus adapter. A new adapter platform came into use in May
2001. The first successful multislit spectroscopic observations
were carried out in September 2003; in the summer of 2004, the
spectropolarimetric mode was implemented, and the first
observations were performed. In the next section, we consider the
optomechanical layout of SCORPIO and its basic characteristics.
Subsequently, we consider the peculiarities of observations in
various modes; these are illustrated using specific results
obtained with the 6-m telescope when the instrument was tested.
In the last section we consider prospects for further upgrading
\SCORPIO.

\begin{table}[t]
\caption{The main characteristics  of  \SCORPIO } \label{tab_main}
\begin{center}
\begin{tabular}{|ll|}
\hline
Total focal ration & $F/2.6$ \\
\hline
\multicolumn{2}{|c|}{Field of view:} \\
full  & $6.1'\times6.1'$ \\
in mutlsilit mode & $2.9'\times5.9'$ \\
\hline
Image scale  & $0.18''/\mbox{px}$ \\
Spectral range & $3\,600-10\,000$\AA \\
\hline
\multicolumn{2}{|c|}{Spectral resolution}\\
with grisms  & \\
(for slit width $1''$) & $1.5-20$\AA \\
with  Fabry-Perot & \\
interferometers & $0.8-2.5$\AA \\
\hline
\multicolumn{2}{|c|}{Maximal quantum efficiency }\\
\multicolumn{2}{|c|}{(telescope+\SCORPIO+CCD)}\\
Direct imaging & 70\% \\
Spectroscopy & 40\% \\
Observations with FPI  & 20\% \\
\hline
\end{tabular}
\end{center}
\end{table}

\section{Description of the spectrograph}
\label{sec1}

Constructionally, \SCORPIO consists of three parts each of which can be used
independently: a focal reducer, a prime-focus adapter platform, and a CCD
detector. Basic parameters of the instrument are given in Table 1.

\subsection{The  focal reducer}

The optical layout of the focal reducer (Fig.1) includes a field
lens and a collimator -- a four-lens apochromat (F/2.2) that
forms the exit pupil of the system, a camera objective -- a
six-lens apochromat (F/1.8), and replaceable optical elements --
FPI, diffraction gratings, filters, a polarization analyzer,
phase plates ant etc. The equivalent focal ratio of the system at
the prime focus of the 6-m telescope is F/2.6. The optical
surfaces are coated with seven antireflecting layers\footnote{The
antireecting layers were put on the surfaces at the Nizhnii
Novgorod Institute of Applied Physics.}  that work in the
wavelength range $3500-10\,000\AA$. The laboratory measurements of
the SCORPIO spectral transmission curve are shown in Fig.2.

Since the collimator optics corrects the coma and field curvature of the
primary mirror of the telescope, we can abandon the use of the standard lens
corrector without antireecting coating. The diameter of the collimated beam is
40 mm. The working focal length of the camera is 14 mm. The equivalent focal
length of the BTA reducer is 15.6 m, which corresponds to an image scale of
75$\mu m/''$. The linear size of the nonvignetted field of view is
$28\times28$ mm in the plane of the detector. Constructionally, the reducer
was made in the form of separate remotely controlled units mounted in a single
case:

\begin{itemize}
\item a multislit unit placed in front of the focal plane;

\item two rotating wheels,each with six positions;

\item a polarization analyzer placed in front of the collimator;

\item the collimator focusing mechanism;

\item the mechanism of putting/withdrawing a dispersive element in/from the
collimated beam.
\end{itemize}

The multislit unit, which is designed for multiobject spectroscopy, is an
arrangement that consists of 16 metal strips with slits located in the focal
plane and moved in a $2\farcm9\times5\farcm9$ field. The slit height is about
18$''$. The position of each slit is fixed using two (holding and catching)
electromagnets. The holding magnets are fixed; the catching magnets are
fastened to the frame that is moved along one coordinate by a stepping motor. A
separate electromagnet simultaneously fixes the positions of all slits in the
focal plane. The arrangement is put in the beam using a stepping motor.

The wheels installed in the spectrograph are designed to put various
replaceable elements in the beam -- filters, slits, masks, etc. All of the
elements installed in the wheel are mounted in bayonet-type holder with a clear
aperture of 72 mm, which allows them to be replaced on the fly. Medium- and
narrow-band interference filters as well as a slit for spectroscopic
observations are generally installed in the wheel located in the focal plane
of the telescope. Broad-band glass filters, a phase plate, and a mask for
slitless spectroscopy are installed in the second wheel (behind the field lens
and in front of the collimator). One position in each wheel always remains
free in order that all of the installed filters could be used in observations.

A 14-mm-thick Savart plate that separates the beams in two
mutually perpendicular planes of polarization by $9''$ in the
focal plane is used as the polarization analyzer. The analyzer
can be turned around the optical axis through $45^\circ$.A
stepping motor is used to put the analyzer in and withdraw from
the beam and to turn it.

The collimator focusing mechanism provides a linear displacement of the
collimator within 12 mm with an accuracy of 0.01 mm.

A slide with two switchable positions to put dispersive elements (FPI or
direct-vision grisms) in the beam is located between the collimator and the
camera. A built-in neon lamp is used to visually adjust the FPI. A central
electromagnetic shutter with a shutter cycle of 0.1 s is located at the flange
of the instrument closest to the primary mirror in front of the mutislit unit.

\begin{figure*}
\centerline{\includegraphics[width=14 cm]{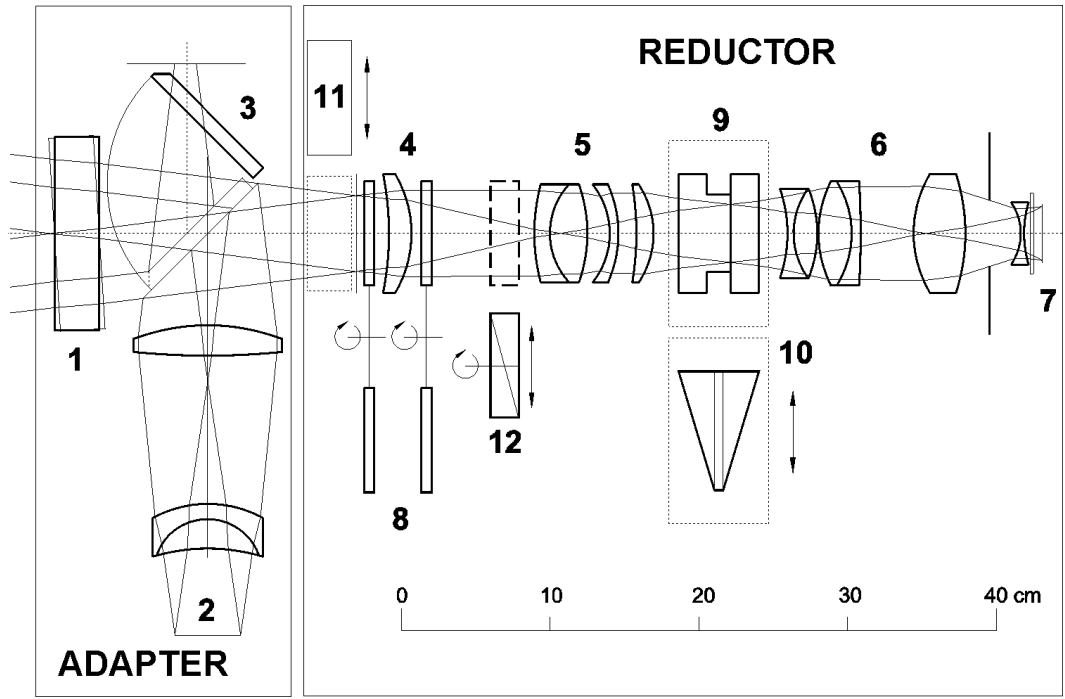}}
\protect\caption{ Optical layout of  \SCORPIO.  (1) -- tip-tilt
plate, (2) -- calibration optics, (3) -- flat mirror, (4) --
field lens, (5) -- collimator, (6) -- camera, (7) -- CCD, (8) --
filter wheels, (9) -- FPI,  (10) --  grism, (11) -- multislit
unit,  (12) -- polarization analyzer.} \label{fig_optics}
\end{figure*}

\begin{figure*}
\centerline{\includegraphics[width=14 cm]{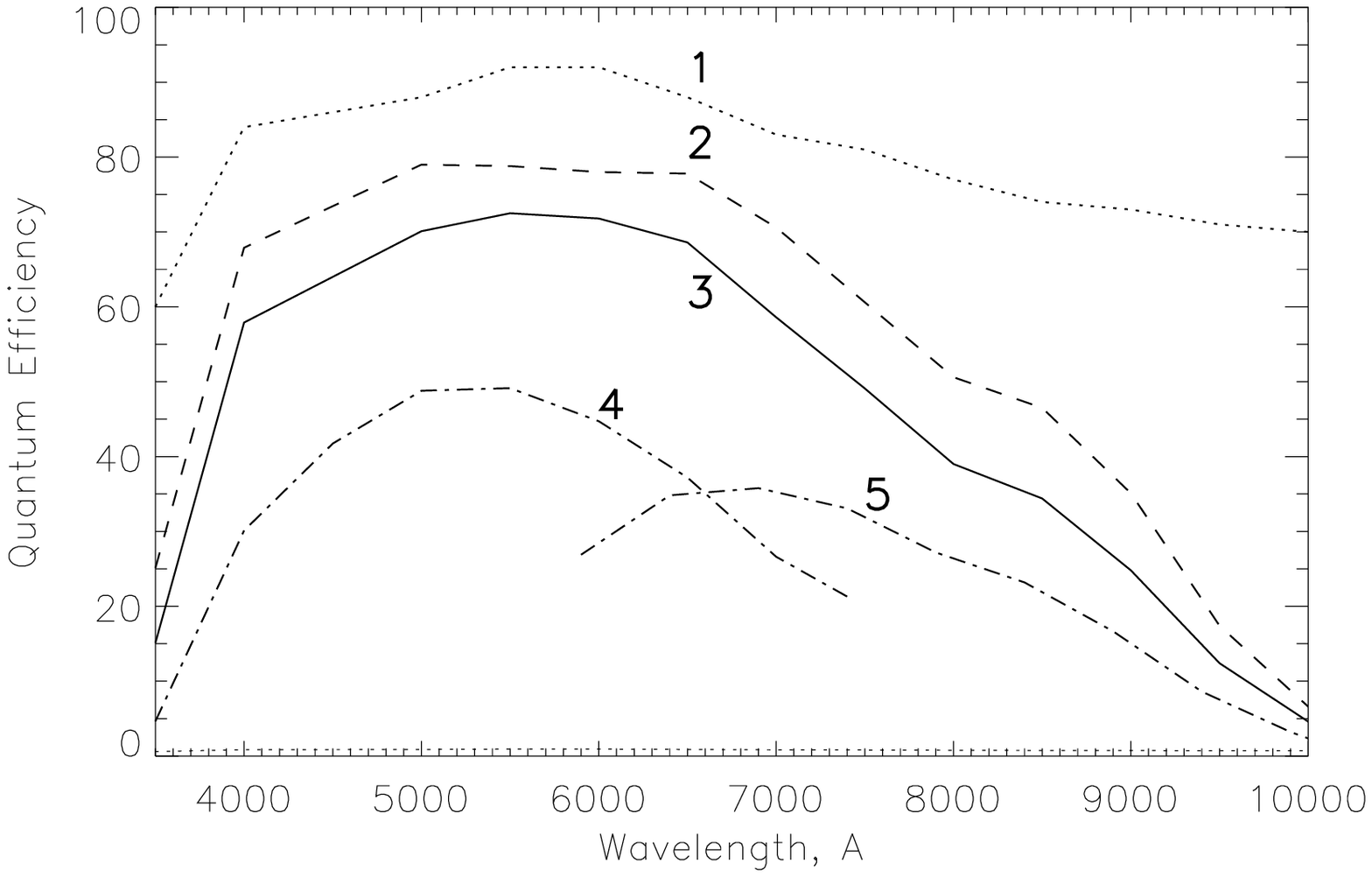}} \protect\caption{
Transmission curve for the \SCORPIO optics (1); the quantum efficiency curve
for the EEV 42-40 CCD, as provided by the producer (2); the combined quantum
efficiency curve for \SCORPIO$+$CCD (3); and the quantum efficiency curve in a
spectral mode for observations with VPHG550G (4) and VPHG550R (5)
low-resolution grisms. }
\end{figure*}

\subsection{The platform-adapter}

The platform adapter is fixed on a turning table in the BTA prime-focus cage and is used for
guiding based on off-axis stars and for illuminating the spectrograph by
calibration lamps. Both the focal reducer and other equipment can be mounted on
it. The adapter is equipped with an electromagnetic shutter that works
independently of the central shutter of the focal reducer.

The adapter contains two rectangular fields to search for guiding stars\footnote{
Since the 6-m telescope has an altazimuth mounting, both the position of the
telescope in $A$ and $z$ and the rotation of the field of view should be
controlled during the guiding. Therefore, two guiding stars are used.}; the centers of the
fields are offset by $12'$ from the center of the field of view. A fiber bundle
displaced by step motors in a rectangular coordinate system is located in
each of the fields. The off-axis lens correction placed in front of each guiding
field corrects the coma of the telescope's primary mirror. The fiber
displacement range is $8\farcm5 \times 4\farcm5$, and the diameter of the field of view of
each fiber is about $40''$.

The flat diagonal mirror (denoted by number 3 in Fig.1) has two fixed positions. At
one of these positions, the mirror throws the images
from the fiber bundles to the TV view. This mode is used when exposing objects.
At the other position, the mirror blocks the central beam of light from the
telescope and throws the image of the field center of the instrument to the TV
view, which is needed to roughly point the telescope at the required object. In
addition, the light from the calibration lamps is thrown to the spectrograph
at this position of the mirror.

The adapter contains the calibration illuminator optics that
forms a convergent beam with a focal ratio of F/4 at the entrance
of the focal reducer, which is telecentrically equivalent to the
beam formed by the primary mirror of the telescope. This scheme
of the calibration unit forms the system's pupil at the same
position where the image of the telescope's main mirror is
located. This allows us not only to properly calibrate the
wavelength scale using a line-spectrum lamp, but also to
calibrate the system's transmission  in various operating modes
(flat field). The entrance area of the calibration unit is
illuminated through an integrating sphere (Ulbricht's sphere) by
two calibration lamps: a He-Ne-Ar-filled lamp to calibrate the
wavelength scale and a continuum halogen lamp to produce a flat
field.

The \SCORPIO  control system contains a relatively large number of
various electromechanical mechanisms (13 motors, two shutters, two
calibration lamps, three crossillumination LEDs, 33
electromagnets in the multislit unit) that are controlled as
follows. Each of the listed units (the focal reducer, the
platform, and the multislit unit) includes electronic boards with
a microprocessor that controls the mechanisms of the corresponding
unit. Instructions to the microprocessor are issued from a remote
personal computer using the standard RS232 data connection
protocol. This implementation of the instrument's remote control
allows the required stability to be achieved when working with
the existing long communication lines of the 6-m telescope. No
continuous control of the motors and their state polling are
required: the microprocessor takes over these functions. At the
same time, any possible changes of the observing technique and
the software on the controlling computer requires no
re-programming of the microprocessors.

\subsection{The CCD detector}

From 2000 until 2003, the detector was a TK1024 $1024\times
1024$-pixel CCD array. Since April 2003, an EEV 42-40
$2048\times2048$-pixel CCD array has been mainly used on \SCORPIO.
Basic parameters of the two detectors are given in Table 2. It should
be noted that a programmed choice of modes with different gain and
readout speeds and noise is possible for EEV 42-40. Figure 2 shows a
plot of the quantum efficiency for EEV 42-40. The detector is cooled
with liquid nitrogen. The entire complex for CCD observations (the
cryostat, the electronics, and the control software) were designed
and produced in the SAO Laboratory of Advanced Design
(http://www.sao.ru/hq/adlab/).

As we see from Table 2, the detectors used have a high quantum
efficiency (see Fig.2), low noise, and low dark current. In addition,
both detectors have an almost perfect surface: the numbers of bad
columns and hot pixels are very small. Perhaps the only shortcoming
is the interference of the transmitted light (fringes) that is
observed at wavelengths longer than 7500 \AA\, for TK1024 and longer
than 6600 \AA\, for EEV 42-40. An appropriate observing technique
(see the section \ref{fringes}) is required to properly subtract the
interference pattern.

\begin{table}
\caption{Parameters of the CCDs} \label{tab_ccd}
\begin{center}
\begin{tabular}{|l|l|l|}
\hline
 &\multicolumn{2}{|c|}{The detector}\\
 \cline{2-3}
         & TK1024 &  EEV-42-40  \\
\hline
 Type     & \multicolumn{2}{c|}{Thin, back-illuminated}\\
\hline
Size & $1024\times1024$& $2048\times2048$ \\
\hline
Pixel size, $\mu$ & 24& 13.5 \\
\hline
Scale$^1$, $''/\mbox{px}$ & 0.32 & 0.18\\
\hline
Field of view$^1$, arcmin & 5.4& 6.1 \\
\hline
Max DQE, \% & 80& 83\\
\hline
Readout noise, \={e}& 3 & 1.8-4\\
\hline
Dark current, \={e}/min & 0.1& 0.03\\
\hline
\multicolumn{3}{l}{$^1$ Till September 2003 the optics supported the focal } \\
\multicolumn{3}{l}{ratio $F/2.9$ was used. In this case the scale on CCD}\\
\multicolumn{3}{l}{TK1024 was $0.28''$/px, with the field of view
$4.8'$}\\
 \hline
\end{tabular}
\end{center}
\end{table}

\begin{table}
\caption{The limiting  magnitudes in broad-band filters (Fatkhullin,
2002) } \label{tab_mag}
\begin{center}
\begin{tabular}{|ccc|}
\hline
Filter  & $T_{exp}$, сек & mags \\
\hline
B       & 2500           & 27.0$^m$ \\
V       & 1500           & 26.3$^m$ \\
R$_C$   & 1260           & 26.4$^m$ \\
I$_C$   & 1800           & 25.1$^m$ \\
\hline
\end{tabular}
\end{center}
\end{table}

\section{Observations in various modes}

\subsection{Direct Images}

SCORPIO is equipped with several filter sets that can be used for
photometric observations. Broad-band glass filters allow the
Johnson-Cousins photometric UBVR$_c$I$_c$ system to be implemented in
direct imaging mode (see Bessell 1990). Table 3 gives the limiting
magnitudes for the detection of faint objects at a signal-to-noise
ratio of 3 at  $1.3''$ seeing. This table is based on the work of
Fatkhullin (2002), who studied the capabilities of \SCORPIO (with a
TK1024 CCD) for the photometry of faint starlike and extended
objects. The set of medium-band interference filters with a bandwidth
of 160-400 \AA\, and central wavelengths of 3700-9700 \AA\, was
produced at the Research Institute of Applied Instrument Making
(Moscow). In direct imaging mode, these filters can be used for
various tasks, such as constructing the spectral energy distribution
for faint objects in the field or imaging extended objects in various
emission lines ($H_\alpha$, [OIII], etc.) and in the continuum.
Examples of such images obtained with \SCORPIO are given in the paper
by Lozinskaya et al. (2002). I.D. Karachentsev (SAO) provided a
filter with a bandwidth of 75 \AA\, centered at the wavelength of the
$H_\alpha$ line, which is used to map the distribution of ionized
hydrogen in nebulae and nearby galaxies. The main problem of
photometric observations is the interference pattern (fringes) during
observations in red photometric bands (see the section
\ref{fringes}). Thus, for example, during I$_c$-band observations
with an EEV 42-40 CCD, the fringes level reaches 8\% of the sky
background level.

\subsection{Long-Slit and Slitless Spectroscopy}

The preimaging possibility proves to be very useful in slit
spectroscopy of both extended objects (since the slit position is
known exactly) and starlike objects if the latter are too faint to be
visible  on the TV-view. Thus, for example, 1-2 minutes  trial
exposure in the filter R$_c$ is enough for reliable pointing to
objects of 22-23$^m$ at a moderate seeing. In the pointing process,
the object under study is set on the detector where the slit image is
projected. Subsequently, the slit is set in place of the filter, and
a grism (the combination of a transparent grating and two prisms)  is
inserted in the collimated beam, which turns the focal reducer into a
fast spectrograph. So, the change of the ``direct imaging''-``long
slit'' configurations takes about one minute.

Figure 3 sequentially shows the process of obtaining observational
data using the spectroscopy of the radio galaxy RCJ 1154+0431 (the
observations at the request of Yu.N. Pariiskii) as an example. The
total V magnitude is $19.8^m$, its measured redshift is z=1.0 (see
Afanasiev et al., 2003a).

SCORPIO is equipped  with set of grisms ensuring observations with
different spectral resolution (from $1.5$ to $20$\AA\, with the slit
width $1''$) in different regions of an optical spectrum. Before
2003, observations were performed using transparent gratings with
profiled grooves with the number of grooves from 300 to 1200 per mm.
The gratings are replicas from cut gratings and were produced at the
S.I. Vavilov State Optical Institute (St.Petersburg). The maximum
quantum efficiency of the entire system (telescope$+$SCORPIO$+$CCD)
with such gratings was 30\% for low-resolution
($\delta\lambda=15-20$\AA) spectra and only about 3-5\% for
higher-resolution ($\delta\lambda=5-6$\AA) spectra. Observations with
grisms using volume phase holographic gratings (VPHGs) were begun in
2003-2004; these have a high transmission and a low level of
scattered light (Barden et al. 2000; Habraken et al. 2001). The
number of lines for the available VPHGs ranges from 550 to 3000 per
mm; in this case, a quantum efficiency of 20-50\% is achieved with
both low (see Fig.2) and high resolutions.

\begin{figure*}
\centerline{\includegraphics[width=14 cm]{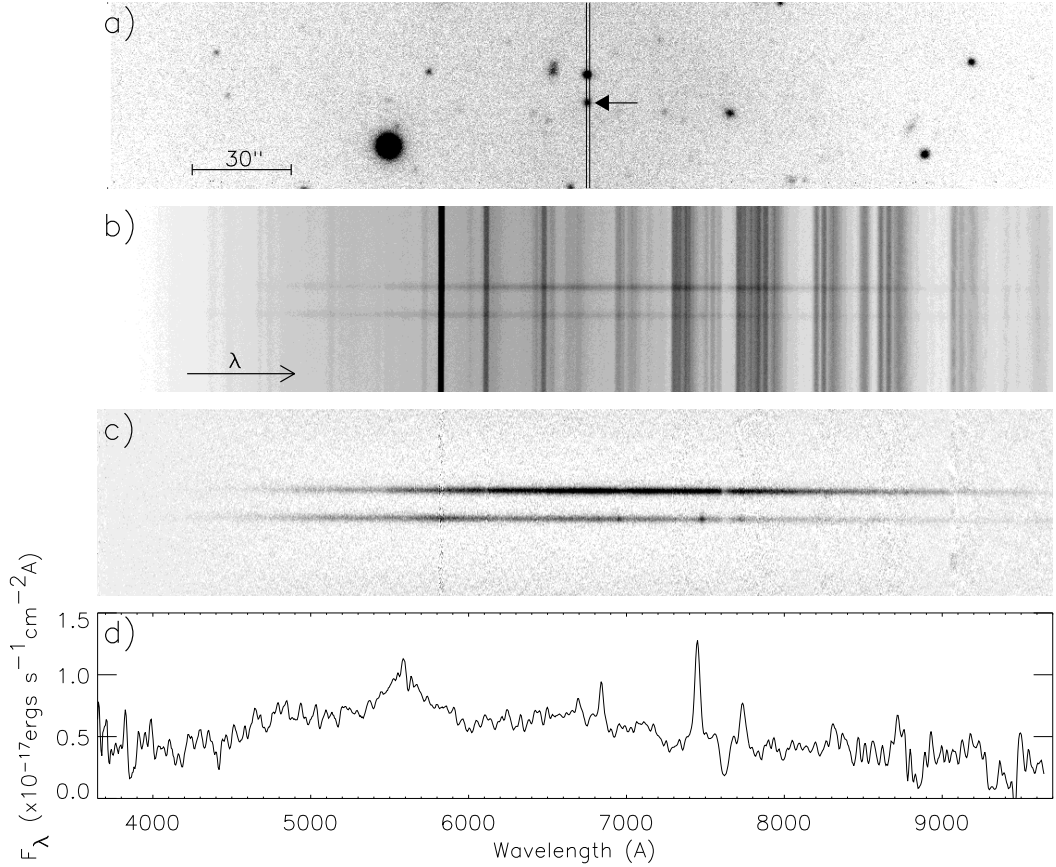}}
\protect\caption{Spectroscopy for the radio galaxy RCJ 1154+0431
with SCORPIO: (a) a fragment of an R$_C$ ($T_{exp}=60$ s) image,
the position of the spectrograph slit is shown; the radio galaxy
is marked by the arrow; (b) a low-resolution spectrum (the sum of
two 600-s exposures); (c) the same after the subtraction of the
night-sky spectrum; (d) the integrated spectrum on the wavelength
scale.} \label{fig_slit}
\end{figure*}

\begin{figure*}
\centerline{\includegraphics[width=18 cm]{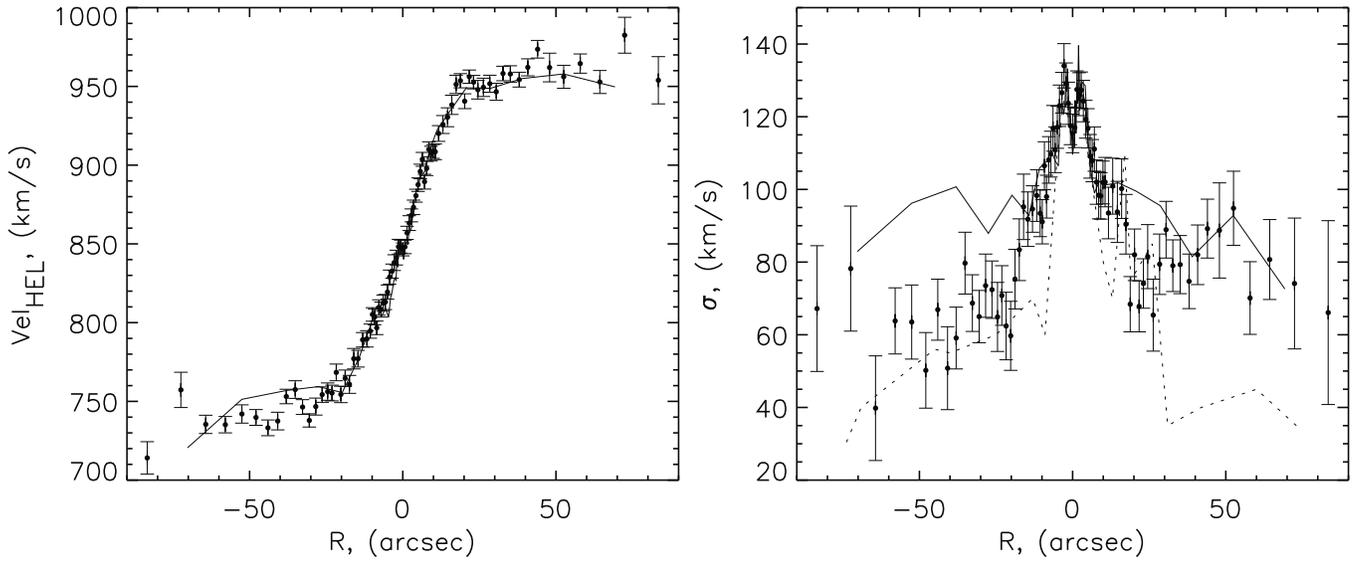}}
\protect\caption{Kinematics of the stellar component in the galaxy
NGC~3412: the distribution of radial velocities (a) and
radial-velocity dispersion (b) along the major axis. The solid and
dashed lines represent the published measurements by Aguerri et al.
(2003) and Neistein et al. (1999), respectively. } \label{fig_stars}
\end{figure*}

\begin{figure*}
\centerline{\includegraphics[width=16 cm]{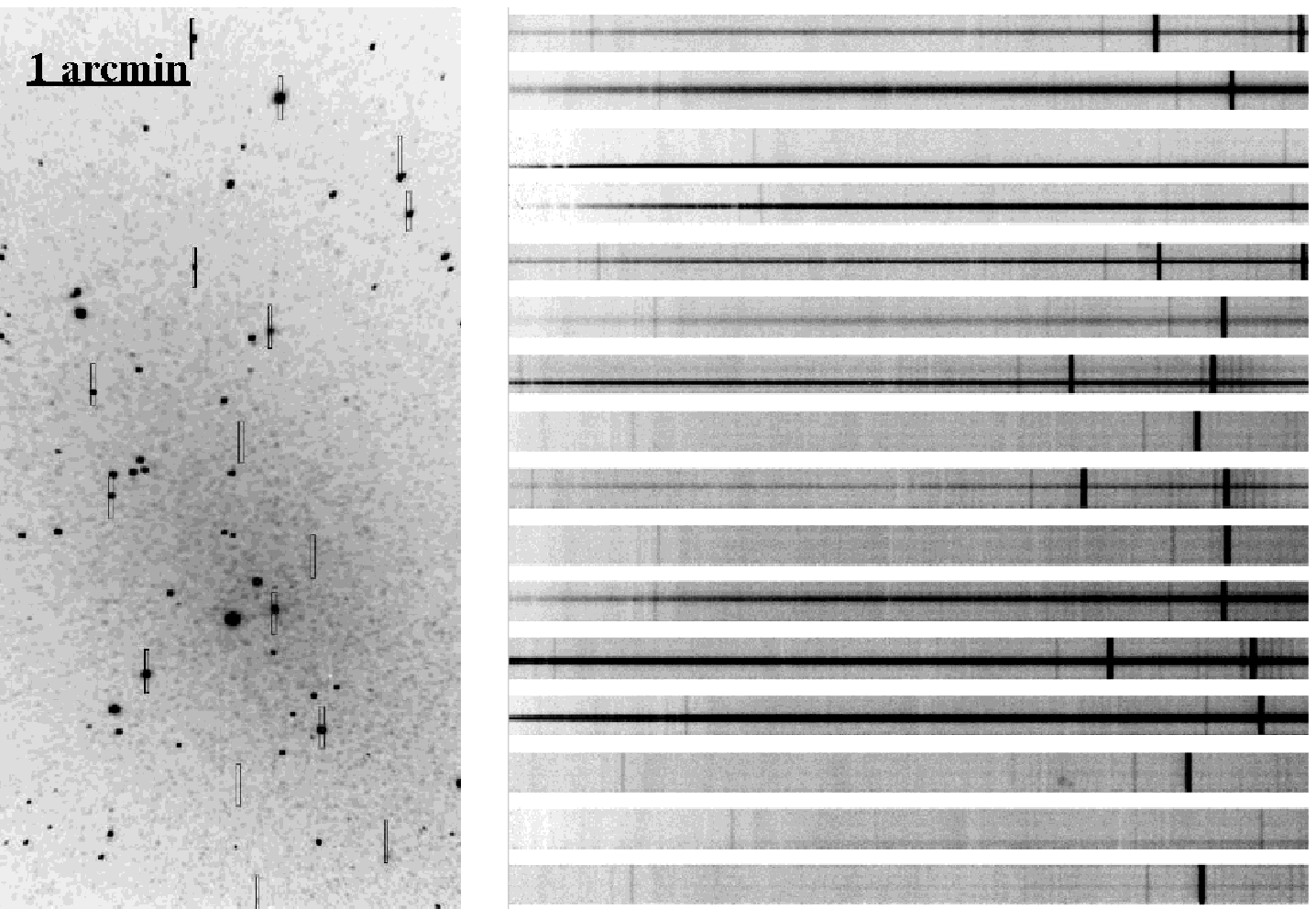}}
\protect\caption{Spectroscopy for globular clusters in the nearby
galaxy NGC~147 (the observations at the request of M.E. Sharina): A
V-band image of the galaxy with marked slit positions (a) and
multislit spectra of the objects (b). } \label{fig_shar}
\end{figure*}

The main set of gratings was produced by Wasath Photonics (USA,
http://wasatchphotonics.com); two gratings were kindly provided by
the University of Padova (Italy).

The achieved quantum efficiency of the instrument has allowed one
to continuously determine the redshifts and spectral
classification of extragalactic radio sources, since the required
low-resolution spectra of $19-21^m$ objects can be obtained even
at moderate atmospheric transparency and at $3-5''$ seeing; the
total exposure time is only 10-20 min (see, e.g., Afanasiev et
al., 2003b; Amirkhanyan et al. 2004). At the same time, at
$1.5''$ seeing in low-resolution spectroscopy of starlike
objects, a limiting magnitude of R$_c=24^m$ is achieved over a
two-hour exposure time (the signal-to-noise ratio is 10 in the
continuum of the spectra obtained). The stability of the
instrumental profile of the spectrograph, which affects both the
accuracy of subtracting the night-sky lines and the possibility
of allowance for the interference pattern in the detector
material (see the section \ref{fringes}), plays a crucial role in
obtaining the spectra of such faint objects. The technique of
displacing an object along the slit between exposures helps
greatly in such observations. During the subsequent reduction, a
pure spectrum, i.e., the spectrum of the sky taken from the same
location, but on the displaced frame, is subtracted from the
spectrum of the object.

A good test for the capabilities of the spectrograph is to study
the kinematics of galactic stellar disks, since absorption
spectra with a relatively high signal-to-noise ratio and a
spectral resolution of at least $\delta\lambda=2-4$\AA\,  should
be obtained here for regions with a surface brightness of
21-23$^m/\Box''$. Figure 4 shows an example of measuring the
parameters of the stellar kinematics along the major axis of a
barred lenticular galaxy, NGC 3412. Over a 1.5-h total exposure
time on \SCORPIO (using VPHG2310), we can measure the radial
velocities and the radial-velocity dispersions of stars for
regions with a V-band surface brightness of $23^m/\Box''$.

\begin{figure*}
\centerline{\includegraphics[width=16 cm]{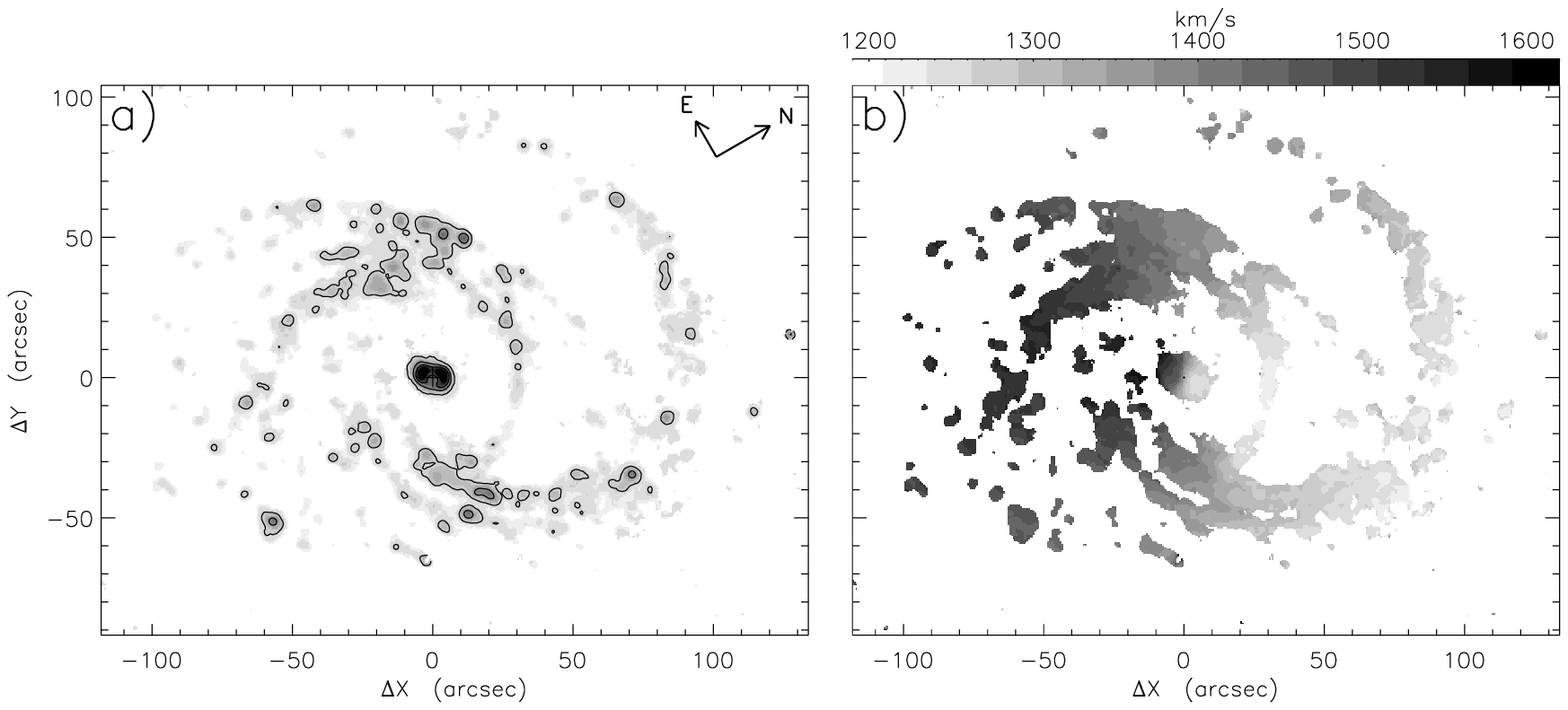}}
\protect\caption{ FPI observations of the galaxy NGC 6951 with
\SCORPIO in the $H_\alpha$: (a) image in the emission line, (b)
the line-of-sight velocity field.} \label{fig_ifp}
\end{figure*}

Slitless spectroscopy, where the slit is replaced with a circular mask about
$30''$ in diameter, is used during the observations of spectrophotometric
standard stars. This technique allows one to completely get away from the
problems of light losses on the slit and distortion of the spectral energy
distribution due to the effect of differential atmospheric refraction.

\subsection{Multislit Spectroscopy}

The available set of grisms can also be used in multislit
observations, although the total spectral range decreases in
comparison with the long-slit case due to the slit displacement in
the field. An IDL-based package of the programs  was written for the
preliminary determination of the optimal position angle of the
multislit unit concerning observed. In such observations, a direct
image of the area under study is obtained, a multislit unit is
introduced in the focal plane of the telescope, and the slits are set
according to the coordinates of the chosen objects measured on the
CCD array. The time of the complete slit arrangement (at the required
accuracy of $0.2-0.3''$) is about 10 min. The spectra obtained are
illustrated in Fig.5.

\subsection{Panoramic Spectroscopy}

The scanning FPI is a highly efficient instrument for studying the
kinematics of extended objects. The observations consist in
sequentially obtaining several tens of images of the interferentic
rings from the object under study (or a calibration lamp) for
various optical paths between the parallel reflecive plates. The
radius of the rings is a function of the wavelength and the FPI
plates' gap. After special reduction, these interferograms can be
represented as a data cube in which two coordinate axes are
located in the plane of the sky and the wavelengths (or the
Doppler velocities measured from the redshifts of spectral lines)
are the third coordinate. In other words, an individual spectrum
is related to each image  pixel.

The Queensgate ET-50 scanning piezoelectric interferometer is
placed between the collimator and the camera where the exit pupil
of the optical system is located. There are two scanning FPIs at
the SAO provided by the Marseilles (France) and Burakan (Armenia)
Observatories. These FPIs work in 235 and 501 orders of
interference (near the $H_\alpha$ line) and provide a spectral
resolution in this line of 2.5 and 0.7\AA\, for spectral ranges
free from order overlapping 28 and 13 \AA\, respectively. These
are successfully used to study both Galactic (nebulae, star
clusters) and extragalactic objects. For more details on the FPI
observations with \SCORPIO, see Moiseev (2002).

Narrow-band filters with a bandwidth of $10-20$\AA\, are used to
separate out the required portion of the spectrum. A set of
filters centered at the wavelength of the redshifted emission
line under study is needed for the observations of various
galaxies. With the filters available at the SAO, objects with
radial velocities from $-200$ to $+10\,000\km s1$ and from +3500
to $+11\,000 \km$ are currently observable in the $H_\alpha$ and
[OIII]$\lambda5007$ lines, respectively. Most of the narrow-band
filters were produced at the Research Institute of Applied
Instrument Making (Moscow); several filters were provided by our
colleagues from the Burakan Observatory (T.Movsesyan) and the
Padova University (G.Barbieri).

Figure 6 shows the FPI observations of the nearby spiral galaxy NGC 6951. The
constructed velocity field is in good agreement with similar observations by
Rozas et al. (2002). Here, the radial velocities were measured with an accuracy
of about $5 \km$. For a detailed discussion of the data obtained, see Moiseev
et al. (2004).

\begin{table*}
\caption{Polarization observations with  SCORPIO } \label{tab_pol}
\begin{center}
\begin{tabular}{|l|r|r|r|r|}
\hline
object  & \multicolumn{2}{|c|}{SCORPIO measurements} & \multicolumn{2}{|c|}{Schmidt et al.(1992)}\\
\hline
        &  P, \%      &$\theta$, $^\circ$         & P, \%      &$\theta$, $^\circ$     \\
\hline
BD+59d389& $6.61\pm 0.11$ & $97 \pm 1$ & $6.701 \pm0.015$ & $ 98.09$ \\
VICyg\#12& $8.80\pm 0.16$ & $117 \pm2$ & $8.947 \pm 0.088$ & $115.03$\\
BD+64d106& $5.35 \pm 0.41$ & $86 \pm 5$ & $5.627 \pm 0.037$ & $96.63$\\
BD+28d4211&$0.33 \pm 0.3$ & --          & $0.054 \pm0.030$  & --\\
\hline
\end{tabular}
\end{center}
\end{table*}

\begin{figure*}
\centerline{\includegraphics[width=16 cm]{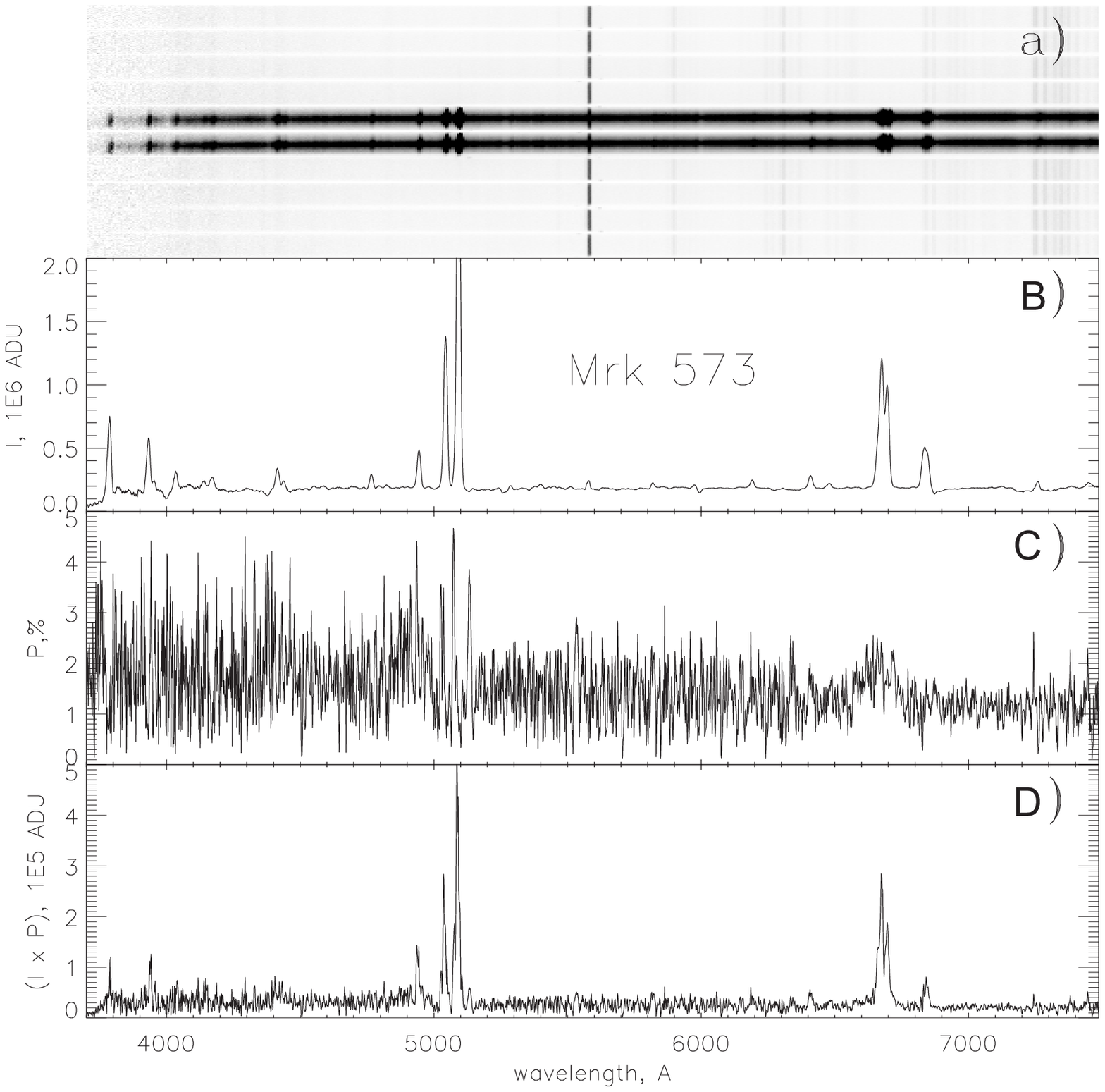}}
\protect\caption{Spectropolarimetry of the Seyfert 2 nucleus in
the galaxy Mrk 573. The grism is VPHG550G, the exposure time is
120 min, and the slit is $1\times7.5''$: (a) the initial spectrum
with the analyzer at position 0; (b) the integrated spectrum of
the nucleus ($V=16^m$) minus the spectrum of the surrounding
galaxy ($I$); (c) the degree of polarization of the nucleus
($P$); (d) the spectrum of polarized emission ($I\times P$).
ADU=0.5\={e}. } \label{fig_pol}
\end{figure*}

\subsection{Polarization observations}

In spectropolarimetric observations, a mask that forms a dotted slit is placed
in front of the \SCORPIO slit. The height of each slit is $7.5''$, and the
separation between the slit centers is $9.8''$. After the introduction of a
polarization analyzer in the beam, a series of pairs of spectra in mutually
perpendicular polarization planes is obtained at the exit of the spectrograph.
Comparing the spectra taken at different orientations of the analyzer, 0 and
$45^\circ$, we unambiguously determine the degree of linear polarization and
the position angle of the polarization plane for the object as a function of
the wavelength. Apart from the count statistics determined by the exposure,
the actual accuracy of measuring the degree of polarization depends on the
level of scattered light in the spectrograph, the accuracy of correcting the
spectra for the flat field, the presence of ghosts in the analyzer (their level
is about 0.2\%), and the accuracy of extracting the spectra from the image.

As our measurements show, the measurement threshold for the
degree of linear polarization in \SCORPIO is 0.2-0.3\%, and the
measurement accuracy is 0.1-0.3\% (depending on the exposure).
Table 4 gives the measurements of polarization standards with
VPHG550G (the dispersion is 2 \AA$/px$) in the V band. The
degrees of polarization that are compared with their published
values were obtained by integrating the spectra in the V band.

Figure 7 illustrates the integrated spectrum of the central region in
Seyfert 2 galaxy Mrk 573 obtained when the \SCORPIO polarization mode
was tested in August 2004. This figure shows the wavelength
dependence of the degree of linear polarization. The spectra were
obtained at $2''$ seeing with VPHG550G. We took a series of 10-min
exposures at successive rotation angles of the analyzer (0 and
$45^\circ$). The total exposure time was 2 h, and the total V-band
brightness of the galactic nucleus in the slit was about $16^m$. A
broad component of the hydrogen lines is distinguished in Fig.7 (c).
The result obtained is in satisfactory agreement with the
observations of other authors (Nagao et al., 2004)

\section{Data Reduction}
\label{fringes}

 The observational data are written in the FITS format. Various standard astronomical
image reduction systems, such as MIDAS or IRAF, can be used for their
reduction. The primary data reduction can be easily automated, since all of
the necessary information about the spectrograph configuration is written in
the FITS-headers. In the IDL environment, we wrote software packages for
SCORPIO data reduction and analysis. The programs for the reduction of FPI,
long-slit, and multiobject spectroscopy have a user-friendly interface and can
be used by users that are not familiar with the IDL language itself. The
reduction sequence of FPI observations with \SCORPIO was described by Moiseev
(2002).

In the reduction of observations, it is important to take into
account the fringe pattern in the sensitive layer of the CCD
array in the red spectral range (see above). In the case of
spectroscopic observations, to correct this effect, the frames
with the object's spectra are divided by the frames with the
accumulated spectra of the built-in lamp, a spectral flat field.
In this way, the fringes  can be reduced by more than an order of
magnitude, which is enough in most cases. We can also get rid of
the fringe pattern by the flat field division in the case of
direct imaging, although using the frames illuminated by the
inner lamp or the twilight sky as a flat field is often not
enough, particularly in the case of broad-band filters, since the
spectral energy distribution of the night-sky background emission
differs markedly from that of the calibration frames. In this
case, an optimal imaging technique is to form an image of the
averaged interference pattern using a series of all accumulations
in the corresponding filter during the observing night.

\section{Further upgrading of the instrument}
\label{sec_modern}

A standard autoguiding system using images on the TV-view (Shergin \&
Maksimova 2001) is used during observations with the 6-m telescope.
It allows the displacement of stars in the field to be compensated by
the corresponding motions of the entire telescope. This is a slow
guiding, since the oscillations of the centers of stellar images at a
frequency of 0.1 Hz are suppressed. In 2005, we planned to put into
operation a fast guiding system based on a tip-tilt fused quartz
plate. This system will allow the oscillations with frequencies up to
10 Hz to be compensated. This will make it possible to significantly
reduce the effect of the telescope's natural oscillations, thereby
improving the quality of stellar images. For an example, using a
local tip-tilt corrector at the Nasmyth-2 focus of the 6-m BTA
telescope would allow the limiting magnitude of the spectrograph
located there to be increased by $0.5-1^m$ (Ivanov et al., 2001).

Equipping the spectrograph with a ``semi-thin''  CCD detector seems
promising. This detector combines the advantages of both ``thick''
(directly illuminated) devices, the absence of a interference
pattern, and ``thin'' devices (back-illuminated) with a high quantum
efficiency. Such a detector has a much smaller  amplitude of fringes
and a higher (up to 80\% at 9000\AA) sensitivity in the red spectral
range, which will allow the SCORPIO efficiency to be increased during
near-infrared observations.

\section{Conclusions}

The high efficiency of the new instrument has been confirmed
during its continuous practical use. Over the period from
September 2000 through November 2004, observations were performed
at the SAO on more than 240 nights; the results obtained were
used in 26 papers, three Ph.D. and two doctoral dissertations.  A
more detailed description of the spectrograph and the observing
technique can be found in the paper by Afanasiev et al. (2004),
while the current description of the spectrograph is accessible
on the Internet at:
\verb*"http://www.sao.ru/hq/moisav/scorpio/scorpio.html", a
continuously updated gallery of observed objects is also given
here. A similar instrument (but without the  platform-adapter and
the multislit unit) produced at the SAO is being successfully
used in observations with the 2.6-m Byurakan Astrophysical
Observatory telescope (Armenia).

\begin{acknowledgements}
We wish to thank the SAO administration for continuous support and
attention when the instrument was designed and produced.  Dodonov
S.N. and  Amirkhanyan V.R. for fruitful discussions in making the
instrument, and  Gazhur E.B.,  Zhelenkov S.R.,  Perepelitsyn E.I.,
and  Fateev V.I., who produced and adjusted the individual parts
of the spectrograph; to the administration of the Institute  of
Astronomy RAS for making available of VPH gratings. This work was
supported in part by the ``Astronomy'' Federal Science and
Technology Program (contract no. 40.022.1.1.1101 from February 1,
2002), the INTAS grant (96-0315) and the Program of the
Department of Physical Sciences of the Russian Academy of
Sciences. And the Program of the Department of Physical Sciences
of the Russian Academy of Sciences. Moiseev A.V. wishes to thanks
the  Russian Science Support Foundation and the Russian
Foundation for Basic Research (project no.04-02-16042) for
partial support of the work.

\end{acknowledgements}

\textit{Translated by V. Astakhov,  V. Shapoval }


\begin{thebibliography}{}
\bibitem{} Afanas'ev V.L.,  Dodonov S.N., Moiseev A.V.,   Verkhodanov O.V.,  Kopylov A.I.,
Pariiskii Yu.N.,   Soboleva N.S.,   Temirova A.V.,   Zhelenkova
O.P.,  Goss W.M.,  2003a,  Astron. Reports,  47,  377
\bibitem{} Afanas'ev V.L.,   Dodonov S.N.,  Moiseev A.V.,  Chavushyan,  V.,  Mujica,  R.,  Juarez,  Y.,  Gorshkov,  A.G.,
Konnikova,  V. K.,  Mingaliev,  M. G.,  2003b,  Astron. Letters,
29,  579
\bibitem{} Afanas'ev V.L.,   Gazhur E.B.,  Zhelenov S.R.,  Moiseev A.V.,  2004,
Bull. SAO,    vol. 58
\bibitem{} Aguerri  J. A.L.,  Debattista V. P.,  Corsini E.M.,
2003,  \mnras,   338.,  465
\bibitem{} Amirkhanyan,  V.R,  Afanas'ev V.L.,  Dodonov S.N.,  Moiseev A.V.,  Mikhailov V.P.,
 2004,   Astron. Letters,   30,  834
\bibitem{} Barden S.C,  Arns J.A.,  Colburn W.S.,  and Williams J.B.,  2000,
\pasp,   112,   809
\bibitem{} Bessell,  M.S.,  1990,  \pasp,   102,   1181
\bibitem{} Buzzoni,  B.,  Delabre B.,  Dekker H.,   et al.,  1984,  ESO Messenger
(ISSN 0722-6691),  Dec. 1984,   9.
\bibitem{} Court\'es,  G.,  1960,  Ann. d'Astrophysics,   23,  115
\bibitem{} Court\'es,  G.,  1964,  \aj,   69,  325
\bibitem{} Fatkhullin T.A.,  2002,  Bull. SAO,   53,   5
\bibitem{} Ivanov A.A.,  Panchuk V.E.,  Shergin V.S.,  2001,  Preprint SAO RAS,  No.155
\bibitem{} Lozinskaya T.A.,   Arkhipova V.P.,   Moiseev A.V.,
Afanasiev V.L.,  2002,   Astron. Reports,   46,  16
\bibitem{} Moiseev A.V.,  2002,  Bull. SAO,   54,  74 (astro-ph/0211104)
\bibitem{} Moiseev A.V.,  Vald\'es J.R.,  Chavushyan V.O.,  2004,  A\&A,  421,  433
\bibitem{} Nagao,  T.,  Kawabata,  K.S.,  Murayama,  T. et al.,  2004,  \aj,   128,   109
\bibitem{} Neistein E.,  Maoz D.,  Rix H.-W.,  Tonry J.L.,  1999,  \aj,  117,   2666
\bibitem{} Nicklas H.,  Seifert W.,  Boehnhardt,  H.,  Kiesewetter-Koebinger,  S.,
Rupprecht,  G.,  1997,  Proc. SPIE,   2871,  in "Optical
Telescopes of Today  and Tomorrow" (Eds. Arne L. Ardeberg),
Nordic Optical Telescope SA (Sweden),   1222
\bibitem{} Rozas M.,  Relano M.,  Zurita A.,  Beckman J.E.,  2002,  A\&A,   386,   42
\bibitem{} Schmidt G.D.,  Elston  R.,  Lupie  O.L.),  1992,  \aj,   104,   1563
\bibitem{}  Shergin V.S.,   Maksimova  V.M.,  2001,  Autoguiding Program TV guide.
User's guide SAO (\verb*"http://www.sao.ru/hq/vsher/vsher_ru.html")
\end{thebibliography}
\end{document}